\documentstyle[aps]{revtex}
\title{Magnetic string contribution to hadron dynamics in QCD}
\author{Yu. A. Simonov}
\address{Institute of Theoretical and Experimental Physics, Moscow, Russia}
\author{J. A. Tjon}
\address{Institute for Theoretical
Physics, University of Utrecht, 3584 CC Utrecht, The Netherlands and
KVI, University of Groningen, 9747 AA Groningen, The Netherlands}
\date{\today}
  \newcommand{\be}{\begin{equation}}
\newcommand{\ee}{\end{equation}} \def\la{\mathrel{\mathpalette\fun
<}} 
\def\fun#1#2{\lower3.6pt\vbox{\baselineskip0pt\lineskip.9pt
\ialign{$\mathsurround=0pt#1\hfil
##\hfil$\crcr#2\crcr\sim\crcr}}}
\newcommand{\vex}{\mbox{\boldmath${\rm x}$}}
\newcommand{\vey}{\mbox{\boldmath${\rm y}$}}

\newcommand{\vep}{\mbox{\boldmath${\rm p}$}}

\newcommand{\vez}{\mbox{\boldmath${\rm z}$}}

\newcommand{\veal}{\mbox{\boldmath${\rm \alpha}$}}

\begin{document}
\maketitle

\begin{abstract}
Dynamics of a light quark in the field of static source
(heavy-light meson) is studied using the nonlinear Dirac equation,
derived recently. Special attention is paid to the  contribution
of magnetic correlators and it is found that it yields a
significant increase of string tension at intermediate distances.
The spectrum of heavy-light mesons is computed with account of
this  contribution and compared to experimental and lattice data.
\end{abstract}
\vspace {0.2cm}

 Pacs: {12.38Gc, 12.38Lg,12.39.Hg}
   \section{Introduction}

   The nature of the QCD string between static charges (the static
   string) was studied  extensively both analytically \cite{1,2,3}
   and on the lattice \cite{4,5,6}. It was shown in these papers that
   the static string is predominantly electric (the connected probing
   plaquette is used for the analysis) and the electric field is
   directed along the string axis.

   In terms of the Field Correlator Formalism (FCM) \cite{7,8} the
   static string is made of the correlators of electric fields only,
   and recent analysis in terms of Casimir scaling \cite{9,10} shows
   that up to 1\% accuracy the string is formed by the bilocal
   correlator of electric fields $D_\Vert^E(x,y) \equiv \langle
   \frac{g^2} {N_c}tr E_\Vert (x) \Phi(x,y) E_\Vert(y)\rangle
  $ (where $\Phi(x,y)$ is the parallel transporter and $E_{\Vert}
  (x)$ is the component of electric field parallel to the string
  axis).  Thus the confinement dynamics for heavy (static) quarks is
  defined by $D^E_\Vert(x,y)$ with this accuracy.

   This fact was used to construct the effective Lagrangian for light
   quarks in the field of the static charge (the heavy--light
   quark-antiquark system) \cite{11,12}. The analysis made in
   \cite{11} has shown that using this Lagrangian one can derive in
   the large $N_c$ limit the nonlinear and nonlocal Dirac equation
   for the light quark Green's function \be (-i\hat \partial -im
   -i\hat M) S=\hat 1 \label{1} \ee where the kernel $\hat M=M(x,y)$
   is proportional to the bilocal (Gaussian) field correlator
$$  \langle \frac{g^2 }{N_c}tr F_{\mu\nu}(x) \Phi (x,y)
F_{\lambda\sigma}(y)\rangle \equiv D_{\mu\nu, \lambda\sigma}(x,y)$$
   and to the Green's function $S$.

   It was shown in \cite{11,12} that the scalar confinement occurs at
   large distances in the selfconsistent solution of (\ref{1}),
   signalling Chiral Symmetry Breaking(CSB). The subsequent analysis
   in \cite{13} has supported this conclusion and the spectrum of the
   heavy-light meson states was calculated together with first estimates
   of the chiral condensate.

In \cite{13} it was assumed that the electric correlators
are dominant in the string and the magnetic part could be neglected.
On the other hand  the analysis of the  heavy quark mass
case (i.e. of equation (\ref{1}) with  the replacement of $M(D,S)$
 by $M(D,S_0)$ where $S_0$ is the free Green's function
for the heavy mass $m$)  done  in \cite{11}, Appendix 5, and
in \cite{14,15}, has shown that the magnetic correlators can also
significantly contribute (at least in the regime when $mT_g\ll 1$,
where $T_g$ is the slope of $D_{\mu\nu, \lambda\sigma}$).  It is  a
purpose of the present paper to study systematically the role of
magnetic correlators for the light quark mass case, $m\ll
\sqrt{\sigma}$, and to make quantitative analysis in this
case. As a byproduct of our analysis the case of heavy quark mass is
reconsidered and some  refinements of previous results are obtained.

\section{Magnetic field contribution to the  confining string}

We study in this section the solution of Eq. (\ref{1}) with the help of
relativistic WKB approach, similarly as in \cite{11}.
The kernel $\hat M$ (where only the bilocal field correlator is
kept and the Gaussian form for it is assumed)
 can be written in the form
  \be
   iM(h,\vex,\vey) = \gamma_\mu
S(x,y)\gamma_\nu J_{\mu\nu} (x,y) \label{2} \ee where \be J_{\mu\nu}
   (x,y) =\exp(-h^2/4T_g^2) J_{\mu\nu} (\vex,\vey) ,~~h=x_4-y_4,
   \label{3}
   \ee
   and
   \be J_{44} \equiv J^{(E)}(\vex, \vey)=\vex\vey
   f_E(\vex,\vey)\frac{\sigma}{2\pi T^2_g},
   \label{4}
   \ee
   \be
   J_{ik} \equiv J_{ik}^{(M)} = (\vex \vey \delta_{ik}- y_ix_k) f_M
   (\vex, \vey)\frac{\sigma}{2\pi T^2_g}.
   \label{5}
   \ee
   Finally,
   \be
   f_{E(M)}(\vex,\vey) =\int^1_0 ds \int^1_0 dt(st)^\alpha \exp
   \left (-\frac{(\vex s-\vey t)^2}{4T^2_g}\right)
   \label{6}
   \ee
   where $\alpha=0$ for $f_E$ and 1 for $f_M$.

   In what follows only asymptotic values of $f_{E, M}$ will be of
   importance, with $|\vex-\vey|\ll |\vex |,|\vey|
\gg T_g$, in which case one has
   \be
   f_E(\vex,\vex) = 3f_M (\vex, \vex)\cong \frac{2\sqrt{\pi}
   T_g}{|\vex |}.
   \label{7}
   \ee

It should be noted at this point, that  sub - and
superscripts $E$ and $M$ refer to the correlators of color-electric
and color-magnetic fields respectively. Due to the structure of Eqs.
(\ref{5}-{6}) for $\vex \cong \vey \rightarrow \infty$ the kernel
is dominated by the color-electric field contribution. On the basis of this,
in \cite{13} the magnetic part of $M$ was disregarded.
However, in what follows we will show that for light quarks the magnetic part
plays an important role at intermediate distances and as a consequence
it can affect strongly the lower lying states of the heavy-light systems.

The kernel $M$, Eq. (\ref{2}), contains the light quark Green's function
$S$, which is a selfconsistent solution of Eq. (\ref{1}).
Following Ref. \cite{11} we can determine $S$ assuming that M
 in the first approximation  is
instantaneous.
In this case $S$ has the spectral decomposition
in terms of eigenvalues $\varepsilon_n$ and eigenfunctions $\psi_n$,
$$ S(h,{\bf x},
{\bf y}) ={i}\left \{ \sum_{n+} e^{-\varepsilon_nh}\theta (h)
\psi_n({\bf x}) \psi_n^+({\bf y})\right.- $$ \be \left. \sum_{n-}
e^ {-\varepsilon_nh}\theta (-h) \psi_n({\bf x}) \psi_n^+({\bf
y})\right \}\gamma_4\equiv i\{\theta(h) S^{(+)}-\theta(-h)
 S^{(-)}\}\gamma_4 \label{3.24} \ee
$\varepsilon_{n^+}=\varepsilon_n^{(+)}$ and $\varepsilon_{n^-}=-
\varepsilon_n^{(-)}$ where $\sum_{n+}$ and $\sum_{n-}$ refer to
sums over positive and negative eigenvalues respectively.

Insertion of (8) in
 (\ref{2}) then yields for $M$
$$
M(h,\vex,\vey)=-i\gamma_\mu S\gamma_\nu J_{\mu\nu}=
$$
\be
\theta(h) [S^{(+)}\gamma_4 J^{(E)}- J^{(M)}_{ik}\gamma_i
S^{(+)}\gamma_4 \gamma_k]-
\theta(-h) [S^{(-)}\gamma_4 J^{(E)}- J^{(M)}_{ik}\gamma_i
S^{(-)}\gamma_4 \gamma_k].
\label{9}
\ee
To find the properties of $\hat M$ we replace $S$ by $S_{lin}$ in
Eq. (\ref{2}), obtaining in this way $M_{lin}$,
where $S_{lin}$ is the quark Green's function of the linear
Dirac equation (\ref{1}) with  the kernel $\hat M$ replaced  by
$\sigma |\vex|\delta^{(4)}(x-y)$ (i.e. the static Dirac
equation with linear potential $\sigma x).$ We now
demonstrate  that $M_{lin}$ indeed tends to
$\sigma|\vex|$ at large distances, and thus yields the a
posteriori proof that the large distance dynamics of the heavy
light quark system is governed by the linear static local
confining potential. At the same time, in the framework of the
same formalism, it will be shown that an intermediate distance
a  region appears, where the dynamics is again local in time and static but with
a larger string tension due to the contribution of magnetic  terms.
For $S^{(\pm)}$ the spherical spinor expansion has the form
$$
S^{(\pm)}
(h, {\bf x}, {\bf y}) =
\sum_{(n^{\pm})}e^{-\varepsilon_n h}
\psi_n({\bf x})
  \psi_n^+({\bf y})=
$$
    \be
  =\sum_{n^{\pm}}\frac{e^{-\varepsilon_n h}}{xy}
  \left (\begin{array}{ll}
  G_n(x)G^*_n(y) \Omega_{jlm} \Omega_{jlm}^*,& -i G_n(x)F^*_n(y)
\Omega_{jlm} \Omega_{jl'm}^{*}\\
  iF_n(x)G^*_n(y) \Omega_{jl'm} \Omega_{jlm}^*,&
  F_n(x)F^*_n(y)\Omega_{jl'm} \Omega_{jl'm}^{*} \end{array} \right )
  \label{3.26}
  \ee
Similarly as in Ref. \cite{11}, we may carry out
   the summation over  partial waves in (\ref{3.26})
   using the WKB approximation for the solutions $G_n, F_n$.
As exploited in \cite{11,16} the results for $S^{(-)}$ can simply
be obtained from $S^{(+)}$ using the symmetry of
$\varepsilon_{n^+}$ and $\varepsilon_{n^-}$ solutions,
namely $\varepsilon_n^{(+)}=\varepsilon_n^{(-)}
\equiv \varepsilon_n$ and
$(\varepsilon_n, G_n, F_n, \kappa)\leftrightarrow
(-\varepsilon_n,F_n, G_n, - \kappa)$.
We quote the final result of the WKB analysis for
   $S^{(\pm)}$, when $|\vex-\vey|\ll |\vex|$
   \be
   S^{(\pm)} = \frac{\sigma e^{-\lambda}}{4\pi y} \delta (1-\cos
   \theta) \left (
   \begin{array}{ll}
   \Delta_1\pm \Delta_0,& X\\
   \bar X&\Delta_1\mp\Delta_0
   \end{array}
   \right)
   \label{11}
   \ee
   where $\lambda=(m+\sigma x)|h|$. The matrix elements $X, \bar X$
   contribute a nongrowing part to $M$ and
   will be of no interest to us in what follows, while $\Delta_1,
   \Delta_0$ are defined as \be \frac{2}{\pi} \int^\infty_1d\tau
   e^{-\lambda (\tau-1)} \frac{\cos
   (a\sqrt{\tau^2-1})}{\sqrt{\tau^2-1}}= \frac{2}{\pi}
   K_0(\sqrt{\lambda^2+a^2}) e^\lambda\equiv \Delta_0(a) \label{3.49}
   \ee and \be \frac{2}{\pi}\int^{\infty}_1 \frac{\tau d\tau
   e^{-\lambda(\tau-1)}}{\sqrt{\tau^2-1}}\cos (a \sqrt{\tau^2-1})
   =\frac{2}{\pi}e^\lambda \frac{\lambda
   K_1(\sqrt{\lambda^2+a^2)}}{\sqrt{\lambda^2+a^2}}\equiv
   \Delta_1(a).
   \label{3.50}
   \ee
   Here $a\cong (\sigma x+m) |x-y|$ for $\sigma x^2\gg 1 $ and $|\vex
   -\vey| \ll x$.
       From expressions (\ref{3.49}) and (\ref{3.50}) we see
   that $\Delta_0, \Delta_1$  are normalized as
    \be \int^\infty_0 \Delta_0(a)
   da = \int^\infty_0 \Delta_1(a) da =1
   \label{3.51}
    \ee
     and  hence  diagonal elements of  $S^{(\pm)}$ behave as smeared
   $\delta$- functions:
   \be
   \int S^{(\pm)} (h, \vex,\vey) d^3 y=\frac12 e^{-\lambda} \left (
   \begin{array}{ll}
   1\pm 1&\\
   &1\mp 1
   \end{array}
   \right).
   \label{15}
   \ee
   Indeed for large $\sigma x$  the functions $\Delta_0, \Delta_1$
   decrease  exponentially fast when $|x-y|$ increases.
   Consider now  Eq. (\ref{1}),
   $$
   (-i\gamma_\mu \partial_\mu-im) S(h,\vex,\vey) -i\int e^{-\frac{
   (h-h')^2}{4T_g^2}} dh' e^{-(\sigma x+m) |h-h'|  }\times
   $$
   $$
   \left\{
   \left(
   \begin{array}{ll}
   \theta(h-h')&\\
   &\theta(h'-h)
   \end{array}
   \right )
   J^{(E)}(\vex,\vex) +
   \left(
   \begin{array}{ll}
   \theta(h'-h)&\\
   &\theta(h-h')
   \end{array}
   \right )
   J^{(M)}_{ik} \gamma_i \gamma_k \right \} S(h',\vex,\vey) =
   $$
   \be
   =\delta^{(4)}(x-y).
   \label{16}
   \ee
In Eq. (\ref{16}) the integration $\int M(\vex,\vez) S(\vez,\vey)
d^3z$ has been carried out using Eq. (\ref{15}). The obtained equation is
a Dirac equation with a time-dependent interaction, localized in
configurational space.
We may now take into account that at large $x\gg T_g$ we have
    \be
    J^{(E)}(\vex,\vex) =\frac{\sigma x}{\sqrt{\pi}T_g};~~
    J^{(M)}_{ik}\gamma_i \gamma_k=\frac23 \frac{\sigma x}{\sqrt{\pi}
    T_g}.  \label{17} \ee

There exist two regions in $\vex$, where Eq. (\ref{16}) can be simplified
further.
Considering the integration over $dh'$ in (\ref{16}), in the
situation when $\sigma x^2\gg 1$ and $T_g\to 0$, we have the two possibilities
\be
i) ~~(m+\sigma x) T_g\ll 1
\label{18}
\ee
\be
ii)~~(m+ \sigma x) T_g\gg 1.
\label{19}
\ee
In the first case, i.e. when (\ref{18}) holds, the leading contribution
 to the integral over $dh'$  come from the region $|h'-h |\la 2 T_g$
because of the first factor in the integrand of Eq. (\ref{16}).
Since the remaining factors vary smoothly over this region,
provided (\ref{18}) is satisfied,
we may replace $h'=h$ in these factors, including $S(h', \vex, \vey)$.
In so doing we get
\be
(-i\gamma_\mu\partial_\mu-im -\frac{i 5\sigma}{3}
x)S=\delta^{(4)}(x-y).
\label{20}
\ee
Note that corrections to the  interaction have the form of a
series in powers of  ($1/(m+\sigma x )T_g)$, and can be neglected in
the first approximation. For this case both color-electric and color-magnetic
terms contribute to the kernel $M$.

Let us now turn to the second case, Eq. (\ref{19}).  Since
$S(h', \vex, \vey)$ varies as $\exp (-(\sigma x+m) |h'|)$
(cf Eq.(\ref{11})) and there is the factor
$\exp (-(\sigma x+m)| h-h'|)$ in (\ref{16}),
it is essential that the above factors being integrated out with
$\theta (h-h')$ or $ \theta (h'-h)$ in the expression (\ref{16}).
For the term with $\theta(h-h')$ we get $\exp(-(\sigma x+m)h)$ and bounds of
integral are defined by $T_g$, whereas for the term with $\theta (h'-h)$
the integration over $dh'$ yields a factor $1/(\sigma x+m)$,
so that this contribution does not grow for large x.
As a consequence the color-magnetic term
can be neglected in case (\ref{19}).

Consequently writing the $S(h,\vex,\vey)$ in the form
\be
S(h, \vex,\vey)=ie^{-(\sigma x+m)|h|}g(\vex, \vey) \left (
\begin{array}{ll}
\theta (h)&\\
&\theta(-h)
\end{array}
\right)
\label{21}
\ee
where $g(\vex,\vey) \approx \tilde \delta^{(3)}(\vex-\vey)$ is a
smeared $\delta$-function, one obtains 
 an equation
\be
(-i\gamma_\mu\partial_\mu-im -{i \sigma} x) S(h,\vex, \vey)=
\delta^{(4)}(x-y),
\label{22}
\ee
where all interaction $\sigma  x$ is due to the electric term
$J^{(E)}(\vex,\vex)$.

In this way we have confirmed a posteriori that solution $S$ of
(\ref{1}) has at large distances the form of the Green's function
for the linear potential, as given by (\ref{22}), and hence our
choice of $S=S_{lin}$ as the first approximation for the kernel $M$,
Eq. (\ref{2}), is justified.

Let us now discuss the regimes (\ref{18}) and (\ref{19}) in more
detail. Consider first the case of heavy quark mass,
 \be
 mT_g\gg 1.
 \label{23}
 \ee
 In this case one automatically obtains the regime (\ref{19}) and
 hence the linear potential as in (\ref{22}). Since $T_g\sim 1
${\rm GeV}$^{-1}$, only top and bottom quarks satisfy (\ref{23}), while
 charmed quark mass lies at the boundary.
 In the limit $m\to \infty$ the Green's function $S_{lin}$
 Eqs. (\ref{3.24},\ref{11}) becomes the standard heavy-quark expression
 \be
 S_{lin} \to S_0= \frac{i e^{-m|h|}}{2} \delta^{(3)}(\vex-\vey) \left
 \{\theta(h) (1+\gamma_4) +\theta (-h)(1-\gamma_4)\right \},
 \label{24}
 \ee
which agrees with Refs.  \cite{11}, \cite{14} and \cite{15}.
However one should not interpret this as corresponding to an admixture of scalar
and vector confining pieces, i.e. $V_{mix}=\sigma x(\frac56+\frac16\gamma_4)$,
since as it was shown explicitly in  \cite{11}, and also here,
in view of the symmetry of the spectrum,
that  in the limiting case $mT_g\gg 1$ the
potential has to be pure scalar $V_{scalar} =\sigma x$, Eq. (\ref{22}).

We now turn to the case $(m+\sigma x) T_g\ll 1$. It is clear that
this condition is valid only for a restricted region  of $x$. However,
for light quarks the selfconsistent solution of (\ref{1}) with the
replacement (\ref{24}) in the kernel $\hat M$ is not
a good approximation for calculating the spectrum of the
lower lying states, since Eq. (\ref{23}) is violated. This conclusion
agrees with the one found in \cite{15}.
For low  mass states, where
the effective region of interaction  $x_{eff}$ satisfies
 \be
 (m+\sigma x_{eff}) T_g\ll 1,
 \label{25}
 \ee
one should use Eq. (\ref{16}) with the approximation  valid for the the regime
$m T_g\ll 1$.
It is a reasonable starting approximation for the whole region of $x$
as long as we are interested for the spectrum, satisfying the
condition (\ref{25}). It should be noted, that in that case the
controversy discussed in Ref. \cite{15} for the replacement
(\ref{24}) does not take place.
Moreover, the  conclusion reached in Ref. \cite{14} that the effective
interaction has the form $V=\frac53 \sigma x$,  applies only to the
states for which $x_{eff}$ satisfies Eq. (\ref{25}),
whereas  for higher excited states  inevitably another regime,
Eq. (\ref{19}), starts to apply with $V_{eff}= \sigma x$ instead of
 $\frac53 \sigma x$.

The region of validity of the magnetic  string tension is important
from the physical point of view, since the additional $\frac23 \sigma
x$ originates from  magnetic field correlators.  From our analysis
we clearly see that at asymptotic large x, i.e. is for very long (and
therefore heavy strings),  the confining mechanism is purely
electric. Moreover, the field contents is independent of the quark masses at the
end of the string. It is only at intermediate regions where the
magnetic contribution may play an important role.
In particular, we have found that
instead of regimes $mT_g\gg 1, m T_g\ll 1$ investigated in
Refs. \cite{11} and \cite{14,15} one has the two
regimes (\ref{18}) or  (\ref{19}) where  the total mass of the string
plus quark mass enters, and  the resulting  confining force is
linear, but with different strength.  For heavy quarks with $m T_g\gg
1$ the regime (\ref{18}) is essentially absent and we may safely use
the color electric confinement mechanism.

 From the phenomenological point of view the lowest states of light
quarks with the property (\ref{15}) feel string tension
$\frac53\sigma$ and this may be important for the resulting masses
of heavy-light systems, such as $D, D_s, B, B_s$ as it will be
 demonstrated in the next section (a similar remark about $D_s$ and
 $B_s$  was made earlier in \cite{14}).

 Moreover, this increase of string tension resolves substantially
(at least for the lowest levels) the descrepancy found for the Regge slopes of
light mesons using relativistic quasi-potential equations for particles with
spin, c.f. \cite{tj}.
In particular, it was found in Ref. \cite{tiem} that a larger string
tension of $\sigma=0.33~ {\rm GeV}^{2}$ than the usually accepted value
of $0.18~{\rm GeV}^2$ was needed to fit the experimental spectrum of the
light mesons. This should be contrasted with the prediction
$\alpha_1=\frac{1}{8\sigma}$ for the Regge slope, given by
$J=\alpha_0+\alpha_1 M^2$, in the case of the spinless Salpeter
equation\cite{lucha}, which is close to the nonrelativistic prediction.
A similar result is found in the case of our nonlocal kernel for the light-heavy
quark system.  Considering the Regge trajectories as obtained  in
Ref. \cite{13} and Table I of this paper  we find for the case of
the light-heavy quark system a value of  $\alpha_1 \cong \frac{1}{\sigma}$.
This is to be compared with the spinless Salpeter prediction for this
system, given by $\alpha_1=\frac{1}{4\sigma}$.
The effect of the covariant treatment of the spin can already be
seen when we use the linear Dirac equation with $V(x)=\sigma x + c_0$ for
heavy-light mesons. In this case we get a slope of $\frac{1}{2\sigma}$ for $c_0=0$.
Most of the difference between our results and those of the linear Dirac
equation can be attributed to having at large distance
an effective negative constant term $c_0$ in the linear potential
present in the case of the nonlocal kernel \cite{13}, which leads to
the shifting of the bound state masses.
It should be noted, that there is another mechanism\cite{sim,wis}
to decrease the Regge slope.
Considering a rotating string\cite{sim} the Regge slope gets somewhat
closer to the nonrelativistic result. One finds a value
of $\frac{1}{\pi\sigma}$ in the case of the light-heavy quark system,
corresponding to a physical half-string.

\section{Spectrum of heavy-light mesons}

The analysis of the previous section suggests that the lowest bound
state solutions of (\ref{1}) can be determined from the
approximate instantaneous nonlocal Dirac equation of the form
\be
(\veal \vep +\beta m) \psi_n(\vex) + \beta \int \tilde M
(\vex,\vez) \psi_n(\vez) d^3 \vez = \varepsilon_n \psi_n (\vex).
\label{26}
\ee
Using the WKB solution for the Green's function $S$ we find that
for small $T_g$ the kernel $\tilde M$ can be approximated by
\be
\tilde M(\vex, \vez) =
\left [ \sqrt{\pi}T_g J^{(E)} (\vex, \vez) +
\int dh' \theta(h') e^{-\frac{(h')^2}{4T_g^2}}
e^{- 2 (\sigma x+m) h'}
J^{(M)}_{ik} (\vex,\vez) \gamma_i \gamma_k\
\right ]
\tilde \delta (\vex,\vez),
\ee
where $\tilde \delta (\vex,\vez)$ is defined as
\be
\tilde \delta (\vex,\vez) = \frac{\sigma}{4\pi z}\delta (1-\cos
\theta_{xz}) [\tilde \Delta_0(a) +\tilde \Delta_1 (a)]
\label{28}
\ee
with $\tilde \Delta_0(a)$ and $\tilde\Delta_1(a)$ denoting the
limiting values of $\Delta_0(a)$ and $ \Delta_1(a)$
respectively when $\lambda\equiv \sigma x h\sim \sigma  x T_g$ tends
to zero.
For low lying states of the light quark system the spatial regions
of interest are expected to satisfy the condition
(\ref{18}), i.e.  $(m+\sigma x) T_g \ll 1$. Hence  we get
\be
\tilde M (\vex, \vez) = \sqrt{\pi} T_g
\{ J^{(E)} (\vex, \vez) + J^{(M)}_{ik} (\vex,\vez) \gamma_i
\gamma_k\} \tilde \delta (\vex,\vez).  \label{27} \ee

It is seen in (\ref{28}) that at large $x$ and $z$ the function
$\tilde \delta (\vex,\vez)$ tends to $\delta^{(3)}(\vex-\vez)$ and
from (\ref{17}) one immediately obtains
\be
\tilde M (\vex, \vez) \cong \frac53 \sigma x \delta^{(3)}(\vex-\vez)
\label{29}
\ee

In what follows we shall solve Eq. (\ref{26}) in two cases $i)$
when $\tilde M(\vex,\vez)$ is replaced by its large distance limit
(\ref{29}) $ii)$ when $\tilde M(\vex,\vez)$ is approximated by
\be
\tilde M(\vex, \vez) = \tilde M_1(\vex, \vez) =\sqrt{\pi} T_g \frac53
J^E (\vex,\vez) \frac{\sigma \delta(1-\cos
\theta_{xz})}{\pi^2\sqrt{xz}}K_0(a).
\label{30}
\ee
The latter approximation is justified in the situation when $|\vex
-\vez|\ll x$, which follows  from the exponential damping of $K_0(a)$
when $|a|$ grows.

One can exploit the computations from \cite{13} to obtain the
eigenvalues $\varepsilon_n (j,l) $ of Eq. (\ref{26})  with the kernel
(\ref{30})  for the lowest states, given in Table 1 for the case of
the vanishing quark mass, $m=0$, and in Table 2 for the mass $m=0.15$
and in Table 3 for $m=0.20$ ${\rm GeV}$.

In what follows we shall  first concentrate on the $s$ -- and
$p$--state eigenvalues and compare them to the results of the QCD sum
rules and lattice calculations. In the language  of heavy quark effective theory
one has the following expansion \cite{17}-\cite{20} for the heavy-light meson
mass $m_H$
\be m_h=m_Q\left (1+\frac{\bar
\Lambda}{m_Q}+\frac{1}{2m^2_{Q^2}}(\lambda_1+d_H\lambda_2) +O\left(
\frac{1}{m^3_q}\right)\right),
\label{31}
\ee
where $\bar \Lambda(n,j,l)=\varepsilon_n(j,l)$. Using
$\sigma=0.18~{\rm GeV}^2$ the solution of (\ref{26})
with the kernel (\ref{30})  yields
the $S$-wave eigenvalue  $\bar \Lambda(0,\frac12,0)\equiv
\bar \Lambda_S$ 
\be
\bar \Lambda_S = 0.520~{\rm GeV}\,\, ~{\rm for}~ \alpha_s=0,
\,\,\,
\bar \Lambda_S= 0.360~{\rm GeV} ~{\rm for}~\alpha_s=0.3,
\label{33}
\ee
which are about a factor of $\sqrt{\frac{3}{5}}$ smaller in the absence
of the magnetic contribution. The predicted values (\ref{33})
should be compared with the results of the QCD heavy-flavour
sum rules \cite{17,18,19,20}
$\bar \Lambda_S=0.57\pm 0.07~{\rm GeV}$
and  the result of the analysis from semileptonic $B$ decays\cite{21}
$\bar \Lambda_S=0.39\pm 0.11~{\rm GeV}$.
A similar value was obtained recently from the QCD sum rules
\cite{22}.
\be
  \bar \Lambda_S=0.45\pm 0.15~{\rm GeV}
\label{36}
\ee
One can see a reasonable  agreement of our results 
with the latest sum rule calculations (\ref{36}). Note here
that we have taken into account the color Coulomb interaction to all
orders, whereas in the QCD sum rules only  the leading  order term
is retained, therefore one may expect that higher orders will
decrease somewhat the value (\ref{36}).

In a similar way one may compute energy eigenvalues for the strange
heavy-light mesons. 
 From Tables 2 and 3 we see that  with $\alpha=0.3$ we have
\be
\bar \Lambda_S^{(s)} =  0.445~{\rm GeV}~{\rm for}~m=0.15~{\rm
GeV},
\,\,\,
\bar \Lambda_S^{(s)} = 0.476~{\rm GeV}~{\rm for}~m=0.20~{\rm
GeV}.
\label{38} 
\ee
These numbers can be compared to the values 
from the experimental $B_s$ and $D_s$ masses. We find
\be
\Delta M_s^{(B)}= M_{B_s}-M_B\cong 90~{\rm MeV},\,\, 
\Delta M_s^{(D)}= M_{D_s}-M_D\cong 100~{\rm MeV}.
\ee
Similar values are found from the spectrum of heavy-light mesons, computed recently on the
lattice \cite{23} for strange mesons. 
 From Eq. (\ref{38}) we see that the experimental data
are close to our predicted value of
$\Delta M_s=\Lambda_S^{(s)}-\Lambda_S= 85~{\rm MeV}$
for $m=0.15~{\rm GeV}$. The various 
available data on  $\Lambda_S$ are summarized in Table 4.

We turn now to orbital and radial excitations.
 For the states with $l=1$, and $j=\frac32$ and $\frac12$ the mass
splitting is due to the spin-orbit interaction inherent in the
Dirac equation.  Denoting these energies as $\varepsilon_n(j,1)\equiv
\bar \Lambda_P(j)$ we find for the nonstrange quark (in GeV):
\begin{eqnarray}
 \bar \Lambda_P(\frac12)&=& 0.817,~~ \bar \Lambda_P
(\frac32)= 0.732~~(\alpha_s=0)
\label{43}
\\
\bar \Lambda_P(\frac12)&=& 0.665,~~ \bar
 \Lambda_P(\frac32) =0.620~~(\alpha_s=0.3).
 \label{44} 
 \end{eqnarray}

Similarly for strange mesons with a strange quark mass 
$m=0.15~{\rm GeV}$, we obtain
\begin{eqnarray}
\bar \Lambda_P^{(s)}(\frac12)&=&0.898,~~ \bar
\Lambda_P^{(s)}(\frac32)=0.832~~(\alpha_s=0) 
\label{45} 
\\
\bar \Lambda_P^{(s)}(\frac12)&=&0.741,~~\bar \Lambda_P(\frac32)=0.712~~
  (\alpha_s=0.3).  
\label{46} 
\end{eqnarray}
These calculations can be compared
with the results of lattice calculations in \cite{23}, with
experiment and the recent QCD sum rule calculations \cite{22}.The
latter  yield for $m=0$
\be 
\bar \Lambda_P=(1\pm0.2)\,{\rm GeV}.
\label{47}
\ee
This value is somewhat higher than the results (\ref{43}) 
and (\ref{44}).
Lattice calculations in [28] give for the difference 
$M(B_J^*)-M(B)\simeq \bar \Lambda_P- \bar \Lambda_S\approx\, 
456~{\rm MeV}$,
which should be compared with our results,
$\Delta \bar \Lambda\equiv \bar \Lambda_P(\frac12)-
\bar \Lambda_S\simeq 305~{\rm MeV}$
for $\alpha_s=0.3$ and with experiment 
$M(B^*_J)-M(B)\simeq 338~{\rm MeV}$. Here $ M(B)=\frac34
M_B(1^-)+\frac14 M_B(0^-)$.
In addition there is a calculation of heavy-light mesons in the
framework of the QCD string approach \cite{24}, where the only
input is current quark masses $(m_n, m_d, m_s),$ string tension
$\sigma$ and $\alpha_s$.  These results have been obtained in
\cite{25} and recently in \cite{26} for real $B, B_s,D,D_s$ mesons
and are easily computed for the limiting case of $m_q\to \infty$,
which yields values listed in Table 5. 
The rather low value found for $\Delta \bar \Lambda$ suggests 
that we still miss some strength in the orbital excitation
in the present work.
In Table 5 the results of different approaches to $\bar
\Lambda_P$ and $\bar \Lambda_P^{(s)}$ are collected. 

Radial excitations are readily obtained from solving Eqs.
(\ref{26}), (\ref{30}) and yield for the $n=1$ state
\be
\varepsilon_1(\frac12,0)=0.951\,\,\,(\alpha_s=0),
\,\,0.805\,\,\, (\alpha_s=0.3)
\label{48}
\ee
and  for the strange meson with $m=150~{\rm MeV}$
\be
\varepsilon_1^{(s)}(\frac12,0)= 1.036\,\,\,(\alpha_s=0),\,\,
0.880\,\,\, (\alpha_s=0.3)
\label{49}
\ee
while for the radial excitation with $l=1$ one obtains
\begin{eqnarray}
&&\varepsilon_1(j,1)= 1.140\,\,\,(\alpha_s=0), \,\,
0.997\,\,\, (\alpha_s=0.3)
\label{50}
\\
&&\varepsilon_1^{(s)}(j,1) =  1.221\,\,\,(\alpha_s=0),\,\,1.076\,\,\,
(\alpha_s=0.3) \label{51}
\end{eqnarray}

These values are compared in Table 6 with the results of the QCD
string approach and recent lattice calculations [28].

  \section{Summary and conclusions}

  The main results of the present paper are of both theoretical and
  phenomenological scope. On the theoretical side it is shown in
  Section 2, that there exist two possible dynamical regimes for the
  quark at the end of the Dirac string,  namely (18) and (19). For the
  case of heavy quark with $mT_g\ll 1 $ only one regime (19) is
  available and results are the same as discussed in [11,14,15]; i.e.
  color magnetic contribution to the string is suppressed in this
  case and one has at large $x$ local Dirac equation with linear
  potential.

  For the case of light quark there is a possibility of another
  regime (18), where color magnetic field also contributes. It is
  shown that this regime operates at intermediate distances and
  yields a static Dirac equation with increased string tension. It is
  demonstrated also that the use in this case in the kernel of
  nonlinear Dirac equation of the full quark propagator $S$ (or its
  WKB approximation $S_{WKB} $) leads to the consistent results,
  while the use of the heavy-mass  propagator $S_0$ - leads
  to inconsistencies shown in [15]. On the phenomenological side the
  regime (18) yields the energy eigenvalues which are in a better
  agreement with other calculations and experiment, as demonstrated
  in Tables 4-6, as compared to our previous results [13] where
  color-magnetic contribution has been neglected.

  The study of the role of color magnetic fields in the dynamics of
  light quarks is at its beginning and the first results call for
  more detailed investigation of the transition between regimes (18)
  and (19) and other applications, e.g. to mesons and baryons
  consisting of light and heavy quarks.

  One of the authors (Yu.S) is acknowledging a partial financial
  support through the RFFI grants 00-02-17836 and 00-15-96785. He is
  also grateful to Yu.S.Kalashnikova and A.V.Nefediev for useful
  discussions and the use of their data prior to publication.

\newpage

\begin{center}
{\bf Table 1}\\

  \vspace{1cm}

\begin{tabular}{|l|l|l|l|l|l|l|l|} \hline
&$\frac12,
0$&$\frac12,1$&$\frac32,1$&$\frac32,2$&$\frac52,2$&$\frac52,3$&$\frac72,3$\\\hline
$\alpha_s=0$&0.520&0.817&0.732&0.934&0.911&1.147&1.070\\
$\alpha_s=0.3$&0.360&0.665&0.620&0.885&0.818&1.057&0.987\\ \hline
\end{tabular} \vspace{1cm}
\end{center}
Ground state energy eigenvalues $\varepsilon_n(j,l)$ in ${\rm GeV}$, $n=0$, 
for two values of $\alpha_s$ and  
$\sigma=0.18~{\rm GeV}^2,~T_g=0.23~{\rm fm}$, m=0.

\begin{center}
{\bf Table 2}\\

  \vspace{1cm}

\begin{tabular}{|l|l|l|l|l|l|l|l|} \hline
&$\frac12,
0$&$\frac12,1$&$\frac32,1$&$\frac32,2$&$\frac52,2$&$\frac52,3$&$\frac72,3$\\\hline
$\alpha_s=0$&0.623&0.898&0.832&1.078&1.010&1.233&1.168\\
$\alpha_s=0.3$&0.445&0.741&0.712&0.965&0.911&1.140&1.081 \\
 \hline \end{tabular} \vspace{1cm}
\end{center}
The same as in Table 1 but for $m=0.15 $ ${\rm GeV}$.

\begin{center}
{\bf Table 3}\\

  \vspace{1cm}

\begin{tabular}{|l|l|l|l|l|l|l|l|} \hline
&$\frac12,
0$&$\frac12,1$&$\frac32,1$&$\frac32,2$&$\frac52,2$&$\frac52,3$&$\frac72,3$\\\hline
$\alpha_s=0$&0.659&0.927&0.867&1.107&1.044&1.263&1.202\\
$\alpha_s=0.3$&0.476&0.769&0.744&0.994&0.944&1.169&1.114 \\
 \hline \end{tabular} \vspace{1cm}
\end{center}
The same as in Table 1 but for $m=0.20 $ ${\rm GeV}$.
\newpage

\begin{center}
{\bf Table 4}\\

  \vspace{1cm}
  Energy eigenvalues $\bar \Lambda_S$ of the heavy-light system in
  the static heavy quark approximation obtained in different
  approaches.\\

\begin{tabular}{|l|l|l|} \hline
Refs.& Method& $\bar \Lambda_S$ (GeV)\\ \hline
24& QCD sum rules& 0.5\\
25& QCD sum rules& 0.4$\div$ 0.5\\
27& QCD sum rules& 0.45$\pm$ 0.15\\
26& Experiment & 0.39$\pm$ 0.11\\
13& Nonlin. Dirac  & 0.287\\
this work & Nonlin.+ magnetic    & 0.360\\\hline
  \end{tabular} \vspace{1cm}
\end{center}

\begin{center}
{\bf Table 5}\\

  \vspace{1cm}

The same as in Table 4 but for $\bar \Lambda_P-\bar \Lambda_s$.

\begin{tabular}{|l|l|l|} \hline
Refs.& Method& $\bar \Lambda_P-\bar \Lambda_S$ (GeV)\\ \hline
27& QCD sum rules& 0.55 $\pm$ 0.35\\
31& QCD string& 0.40 \\
28& Lattice & 0.47\\
PDG& Experiment  & 0.383\\
this work & Nonlin.+ magnetic    & 0.305 ($\bar
\Lambda_P(\frac12) -\bar \Lambda_S$)\\\hline \end{tabular}
\vspace{1cm} \end{center}

\begin{center}
{\bf Table 6}\\
\vspace{1cm}
The same as in Table 4, but for the radial excitation,
$\Lambda'_S-\Lambda_S$.

\begin{tabular}{|l|l|l|}\hline
Refs.&Method& $\Lambda'_S-\Lambda_S$ (GeV)\\ \hline
32& experiment, $M_{B^*}-M_B$& 0.581\\
28& lattice, $M_{B^*}-M_B$& 0.602\\
31&QCD string& 0.564\\
this work& Nonlin.+magn.&0.631\\\hline
\end{tabular}
\end{center}

   \end{document}